\documentclass[sigconf]{acmart}
\usepackage{subfig} 
\usepackage{xcolor}
\usepackage{fontawesome} 
\usepackage{pifont}
\usepackage{natbib}
\usepackage{times} 
\usepackage{enumitem} 
\usepackage{url} 
\usepackage{hyperref}
\usepackage{multirow}

\AtBeginDocument{%
  }

\setcopyright{acmcopyright}
\copyrightyear{2025}
\acmYear{2022}
\acmDOI{XXXXXXX.XXXXXXX}

\acmConference[MM '26]{Proceedings of the 33rd ACM International Conference on Multimedia}{November 10--14, 2026}{Rio de Janeiro, Brazil}
\acmISBN{978-1-4503-XXXX-X/2018/06}
\renewcommand\footnotetextcopyrightpermission[1]{}
\begin{document}

\title{AT-ADD: All-Type Audio Deepfake Detection Challenge Evaluation Plan }

\author{Yuankun Xie}
\authornote{These authors contributed equally to this work.\\
Official Website: \url{https://at-add.com}. \\
Organization Inquiries: Haonan Cheng (haonancheng@cuc.edu.cn), \\
Jiayi Zhou (zjy326112@antgroup.com). \\
Technical Questions: Yuankun Xie (xieyuankun@cuc.edu.cn), \\
Tao Wang (mengyu.wt@antgroup.com).}
\affiliation{%
  \institution{Communication University of China \& Ant Group}
  \city{Beijing}
  \country{China}
}

\author{Haonan Cheng}
\authornotemark[1]
\affiliation{%
  \institution{Communication University of China}
  \city{Beijing}
  \country{China}
}

\author{Jiayi Zhou}
\authornotemark[1]
\affiliation{%
  \institution{Machine Intelligence, Ant Group}
  \city{Shanghai}
  \country{China}
}

\author{Xiaoxuan Guo}
\affiliation{%
  \institution{Communication University of China}
  \institution{Ant Group}
  \city{Beijing}
  \country{China}
}

\author{Tao Wang}
\affiliation{%
  \institution{Machine Intelligence, Ant Group}
  \city{Shanghai}
  \country{China}
}

\author{Jian Liu}
\affiliation{%
  \institution{Machine Intelligence, Ant Group}
  \city{Shanghai}
  \country{China}
}

\author{Weiqiang Wang}
\affiliation{%
  \institution{Machine Intelligence, Ant Group}
  \city{Shanghai}
  \country{China}
}

\author{Ruibo Fu}
\affiliation{%
	\institution{Institute of Automation, Chinese Academy of Sciences}	
	\city{Beijing}
	\country{China}
}

\author{Xiaopeng Wang}
\affiliation{%
	\institution{Beijing Institute of Technology}
	\city{Beijing}
	\country{China}
}

\author{Hengyan Huang}
\affiliation{%
	\institution{Communication University of China}
	\city{Beijing}
	\country{China}
}

\author{Xiaoying Huang}
\affiliation{%
	\institution{Communication University of China}
	\city{Beijing}
	\country{China}
}
\author{Long Ye}
\affiliation{%
	\institution{Communication University of China}
	\city{Beijing}
	\country{China}
}

\author{Guangtao Zhai}
\affiliation{%
	\institution{Shanghai Jiao Tong University}
	\city{Shanghai}
	\country{China}
}

\renewcommand{\shortauthors}{Xie et al.}

\begin{abstract}
The rapid advancement of Audio Large Language Models (ALLMs) has enabled cost-effective, high-fidelity generation and manipulation of both speech and non-speech audio, including sound effects, singing voices, and music. While these capabilities foster creativity and content production, they also introduce significant security and trust challenges, as realistic audio deepfakes can now be generated and disseminated at scale. Existing audio deepfake detection (ADD) countermeasures (CMs) and benchmarks, however, remain largely speech-centric, often relying on speech-specific artifacts and exhibiting limited robustness to real-world distortions, as well as restricted generalization to heterogeneous audio types and emerging spoofing techniques.
To address these gaps, we propose the \emph{All-Type Audio Deepfake Detection (AT-ADD)} Grand Challenge for ACM Multimedia 2026, designed to bridge controlled academic evaluation with practical multimedia forensics. AT-ADD comprises two tracks: (1) \emph{Robust Speech Deepfake Detection}, which evaluates detectors under real-world scenarios and against unseen, state-of-the-art speech generation methods; and (2) \emph{All-Type Audio Deepfake Detection}, which extends detection beyond speech to diverse, unknown audio types and promotes type-agnostic generalization across speech, sound, singing, and music. By providing standardized datasets, rigorous evaluation protocols, and reproducible baselines, AT-ADD aims to accelerate the development of robust and generalizable audio forensic technologies, supporting secure communication, reliable media verification, and responsible governance in an era of pervasive synthetic audio.
\end{abstract}


% \ccsdesc[500]{Security and privacy~Social aspects of security and privacy}
% \ccsdesc[500]{Applied computing~Sound and music computing}
% \ccsdesc[500]{Computing methodologies~Artificial intelligence}

\vspace{-3pt}
%  \includegraphics[width=3in]{figure/problem}


\maketitle
\vspace{-3pt}

\section{Introduction}

\begin{figure}[!tb]
	\centering
	\subfloat{\includegraphics[width=3.3in]{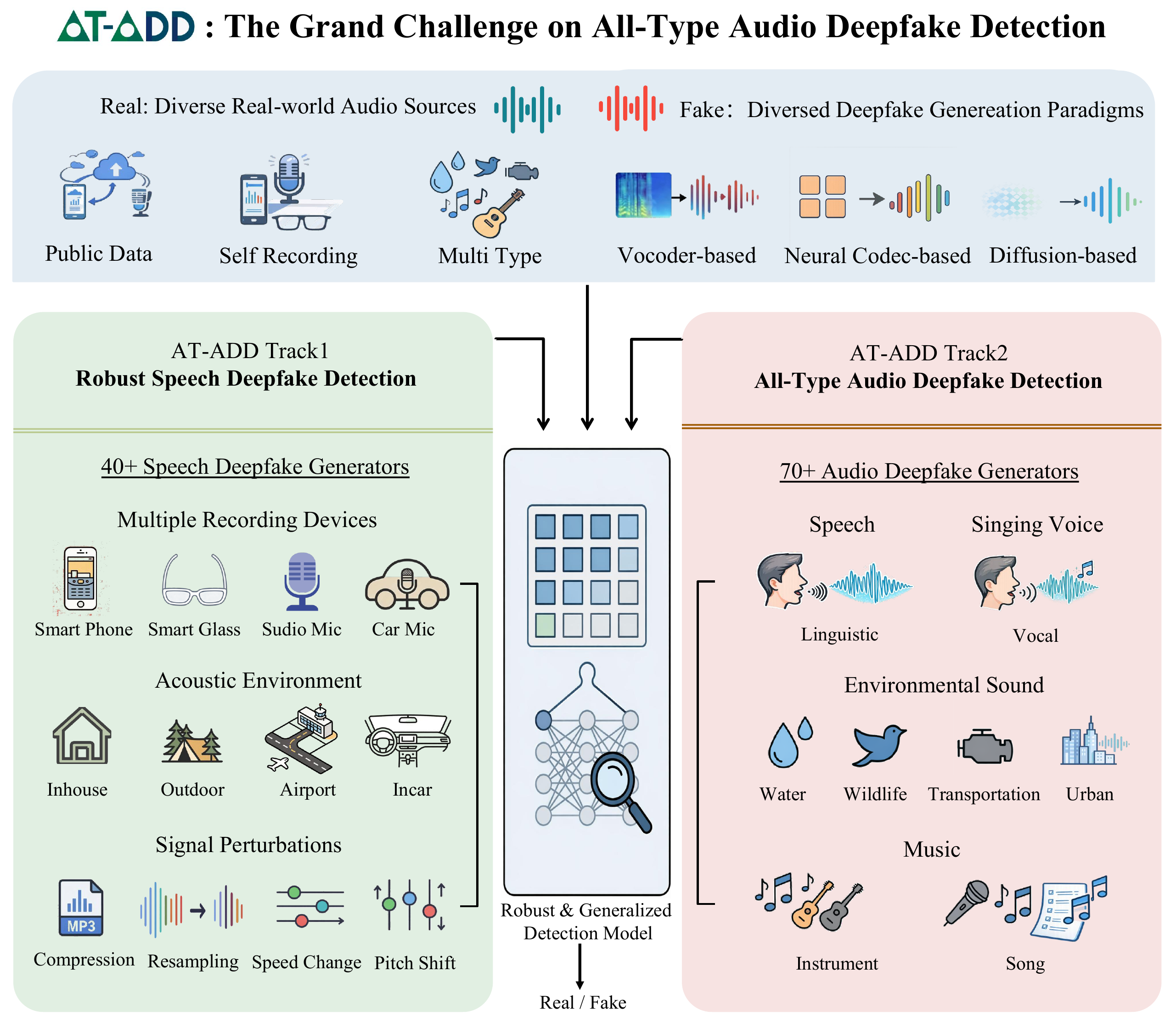}}
	\hfil
	\caption{AT-ADD challenge overview.}
	\label{fig:intro} 
\end{figure}

Recent advances in audio generation technologies, particularly Audio Large Language Models (ALLMs), have significantly improved the realism, scalability, and accessibility of synthetic audio. Modern generative systems are now capable of producing high-fidelity audio across a wide range of content types, including speech, environmental sounds, singing voices, and music. While these developments greatly benefit content creation and multimedia applications, they also introduce serious security and trust risks, as audio deepfakes can be generated and disseminated at scale with increasing realism.

Despite growing research efforts in audio deepfake detection (ADD), existing methods and benchmarks remain largely focused on speech and are typically evaluated under relatively controlled conditions. Consequently, current countermeasures (CMs) often rely on speech-specific artifacts and exhibit limited robustness when deployed in real-world scenarios involving channel variability, environmental noise, compression, replay attack and other distortions. Furthermore, their generalization capability remains insufficient when faced with emerging generation paradigms, such as ALLM-based synthesis and neural codec-driven generation, as well as diverse non-speech audio types.

To address these challenges, we introduce the AT-ADD (All-Type Audio Deepfake Detection) Grand Challenge at ACM Multimedia 2026. The goal of AT-ADD is to bridge the gap between idealized research settings and real-world multimedia forensics by systematically evaluating both robustness under realistic conditions and generalization across audio types and unseen generation methods.

\textbf{Track 1: Robust Speech Deepfake Detection.}
This track focuses on robustness in real-world speech deepfake detection. We construct a large-scale dataset (AT-ADD Track 1) covering more than 40 state-of-the-art speech generation models, spanning vocoder-based, neural codec-based, and diffusion-based paradigms, with particular emphasis on emerging ALLM-driven synthesis. The real speech data are collected from multiple public datasets and in-the-wild recordings, covering diverse languages, recording devices (e.g., smartphones, in-vehicle systems, wearable devices), and acoustic environments. To simulate realistic deployment conditions, we further introduce a wide range of degradations, including background noise, reverberation, replay attack, compression, resampling, and speed perturbation. This track aims to evaluate the robustness and cross-domain generalization of detection systems under complex real-world conditions, encouraging models to move beyond reliance on narrow artifact cues toward learning more fundamental and transferable forensic representations.

\textbf{Track 2: All-Type Audio Deepfake Detection.}
Building upon Track 1, this track extends the detection task from speech to all types of audio, including environmental sounds, singing voices, and music. We construct a new dataset (AT-ADD Track 2) covering more than 70 audio generation models across different generation mechanisms. Unlike Track 1, no additional  signal-level perturbations are introduced in this track, allowing for a more controlled investigation of cross-type generalization. Real audio is collected from multiple public datasets, while synthetic audio is generated under a unified framework without introducing additional distortions. Participants are required to design type-agnostic CMs capable of distinguishing real and fake audio under unseen audio categories. This track encourages models to capture shared synthesis artifacts across different audio types, promoting the development of universal audio deepfake detection methods.

Together, the two tracks form a progressive evaluation framework: Track 1 emphasizes robustness under realistic conditions within the speech domain, while Track 2 focuses on generalization across audio types and domains. This design reflects the evolving landscape of audio deepfake threats—from speech-centric manipulation to open-domain audio generation—and provides a structured benchmark for advancing robust and generalizable audio forensic technologies.

\section{Challenge Tasks}
\label{sec:tasks}

The AT-ADD challenge is organized into two complementary and progressively structured tracks under a \textbf{closed setting}. This design aims to assess which types of CMs can achieve stronger generalization and robustness under limited data conditions. Participants are required to train their CMs strictly using only the data provided by the organizers, thereby ensuring a fair and controlled comparison across methods.

\subsection{Track 1: Robust Speech Deepfake Detection}
\label{sec:track1}

\textbf{Goal.}
Track~1 aims to bridge the gap between existing benchmarks and real-world deployment scenarios for speech deepfake detection. It evaluates whether a detector can remain reliable under realistic domain shifts and practical post-processing effects, while maintaining strong performance against modern high-fidelity synthesis systems.

\textbf{Task definition.}
Given an input speech utterance, participants are required to predict whether the input is \emph{real} or \emph{fake}. In this task, \emph{fake} refers specifically to deepfake speech generated using deep neural network-based methods, while \emph{real} refers to non-deepfake speech. It should be noted that signal distortions or transformations, such as compression, resampling, speed perturbation, and pitch shifting, as well as replay-based attacks, do not change the original real/fake label in this task. The training and development data are fully provided by the organizers, and the use of external data is not allowed under the closed setting.

The evaluation set includes deepfake samples generated by methods that are \emph{unseen} during training and reflect recent state-of-the-art generation techniques. Meanwhile, the real speech in the evaluation set is collected under realistic conditions, involving variations in recording devices, acoustic environments, languages, and other real-world factors. 

\textbf{Long-term research target.}
Track~1 is designed to promote the development of practically deployable CMs that generalize across unseen generators and real-world recording conditions. By combining controlled training data with an evaluation set featuring both unseen deepfake methods and significant domain shifts, this track encourages advances in robustness, generalization, and reliability beyond conventional clean-benchmark settings.

\subsection{Track 2: All-Type Audio Deepfake Detection}
\label{sec:track2}

\textbf{Goal.}
Track~2 targets universal audio deepfake detection across heterogeneous audio types and aims to develop \emph{type-agnostic} detectors that generalize across both audio types and unseen generation methods.

\textbf{Task definition.}
Given an input audio clip of unknown type, participants are required to determine whether it is \emph{real} or \emph{fake}. In this task, \emph{fake} denotes deepfake audio generated by deep neural network-based methods, whereas \emph{real} denotes non-deepfake audio. Notably, in Track~2, audio-type labels (i.e., speech, sound, singing, and music) are \emph{not} available at test time, reflecting realistic deployment scenarios.

Similar to Track~1, this track follows a closed setting, where participants must use only the provided training and development data, without access to external resources.

\textbf{Long-term research target.}
Track~2 is designed to move beyond speech-centric audio deepfake detection and promote the development of unified CMs capable of handling both unknown audio types and unseen generation methods. This setting is motivated by real-world scenarios, where the type of incoming audio is often unknown in advance and may extend beyond speech alone. By encouraging the learning of shared and transferable representations across diverse audio types, Track~2 aims to advance universal CMs that are better suited to realistic and heterogeneous application environments.

\section{Related Work}

ADD has progressed rapidly with the rise of high-fidelity neural generative models. Existing studies are dominated by speech-focused benchmarks and methods, where SSL representations and attention-based back-ends have driven substantial performance gains. In contrast, detection for non-speech audio (sound, singing voice, and music) remains less explored and is still largely benchmark-driven, while cross-type generalization across heterogeneous audio domains is an emerging yet under-established research frontier. In the following, we review prior work by audio type---speech, sound, singing voice, and music---and then discuss cross-type ADD, which motivates the need for a unified and realistic all-type benchmark.

\textbf{Speech}. Speech deepfake detection has been extensively studied, largely driven by the ASVspoof challenges \cite{nautsch2021asvspoof, liu2023asvspoof, wang2024asvspoof} and ADD challenges \cite{yi2022add, yi2023add}. Representative CMs include AASIST \cite{jung2022aasist} and SSL-based pipelines that combine XLSR with AASIST \cite{tak2022automatic}. Subsequent studies have investigated different SSL representations \cite{phukan2024heterogeneity,kheir2025comprehensive}, layer utilization of SSL features \cite{zhang2024audio,wang2025mixture,pan2024attentive}, and robustness \cite{zhang2025i,xu2025alden,kawa2023defense}. However, a substantial gap remains between existing public benchmarks and real-world conditions (e.g., diverse capture devices, channel effects, and replay attack \cite{muller2025replay, zhang2025echofake}), highlighting the need for new datasets and protocols that better reflect realistic deployment scenarios and enable more faithful evaluation of detection performance in the wild.

\textbf{Sound}. Compared to speech deepfake detection, research on  environmental sound deepfake detection is still relatively nascent and is largely driven by dataset and benchmark construction. The Environmental Sound Deepfake Detection (ESDD) Challenge \cite{yin25_interspeech} has recently advanced this area by covering a wide range of ALLM-based text-to-audio (TTA) and audio-to-audio (ATA) synthesis methods. Current state-of-the-art solutions typically leverage sound-oriented SSL representations such as SSLAM \cite{guo2025envsslam}.

\textbf{Singing voice}. Singing voice can be considered a subcategory of music; however, it is treated as a distinct audio type in this challenge due to its unique characteristics and the high difficulty of deepfake song detection. Unlike general music, singing voice shares strong similarities with speech as both are produced by human vocal mechanisms, while also exhibiting complex musical structures such as melody and rhythm. These properties make singing voice particularly challenging for existing CMs. Recent work, such as SVDD \cite{zhang2024svdd}, has promoted research in deepfake singing voice detection. Competitive approaches often leverage hybrid representations by combining speech-oriented SSL features (e.g., XLSR) with music-oriented SSL models such as MERT \cite{li2024mert} and WavLM \cite{chen2022wavlm}, as explored in recent studies \cite{10832226,zhang2024xwsb,chen2024singing}.

\textbf{Music}. Music represents a broad and diverse audio type that encompasses both instrumental compositions and songs. Compared to speech and singing voice, music exhibits higher variability in structure, timbre, and generation mechanisms, posing additional challenges for deepfake detection.
FakeMusicCaps \cite{comanducci2024fakemusiccaps} provides a benchmark for synthetic-music detection and enables the study of text-to-music (TTM) generation artifacts. However, methodological explorations remain relatively limited compared to speech \cite{li2024detecting,wei2025voices}.

\textbf{Cross-type.} A few studies have investigated transfer across audio types, for example between speech and singing voice \cite{gohari2025audio}, from speech to music \cite{li2024audio}, and ESDD 2 (from sound to both speech and sound deepfake detection settings \cite{zhang2026esdd2}). Xie et al.~\cite{xie2025detect} further establish an SSL-based benchmark for all-type ADD and propose wavelet prompt tuning to improve cross-type generalization. However, these studies are still grounded on relatively limited and task-specific datasets, and the community is still lacking a comprehensive and widely accepted benchmark that systematically covers the full spectrum of audio types and realistic conditions.

Overall, despite substantial progress in speech deepfake detection, a clear gap remains between academic benchmarks and real-world deployment, particularly in terms of robustness to complex acoustic environments, and rapidly evolving deepfake paradigms. This motivates the need for a robust speech deepfake detection benchmark that more faithfully reflects practical scenarios. Meanwhile, research on all-type audio deepfake detection is still at an early stage, and a unified evaluation model spanning speech, sound, singing voice, and music is essential to drive the next generation of generalizable CMs.

\section{Datasets and Resources}

To support the AT-ADD challenge, we construct two benchmark datasets: \textbf{AT-ADD Track~1} and \textbf{AT-ADD Track~2}, corresponding to robust speech deepfake detection and all-type audio deepfake detection, respectively. For both tracks, we provide standardized train, development (dev), and evaluation (eval) splits under a closed setting. In addition, a progress subset is provided for progress evaluation, which is sampled from the evaluation set with the same distribution and constitutes 20\% of the full eval set. An overview of the two tracks is presented in Table~\ref{tab:alldata_protocol}.

Detailed dataset compositions and statistics are presented in the following subsections. It should be noted that, to ensure data quality (e.g., by removing fully silent segments), we applied a series of screening and filtering procedures. As a result, the number of samples in each condition is not perfectly uniform. 
\begin{table*}[t]
\centering
\caption{AT-ADD statistics (number of clips) for Track 1 and Track 2. * indicates withheld statistics.}
\label{tab:alldata_protocol}
\begin{tabular}{c c | ccccc}
\hline
\multirow{2}{*}{\textbf{Split}} & 
\multirow{2}{*}{\textbf{T1}} & 
\multicolumn{5}{c}{\textbf{T2}} \\
\cline{3-7}
 &  & \textbf{Speech} & \textbf{Sound} & \textbf{Singing} & \textbf{Music} & \textbf{Total} \\
\hline
Train    & 49,575  & 49,575  & 39,840 & 36,000 & 21,366 & 146,781 \\
Dev      & 49,734  & 49,734  & 19,929 & 16,000 & 5,406  & 91,069 \\
Progress & 29,269  & *  & *  & *  & *  & 45,875 \\
Eval     & 146,346 & * & * & * & * & 229,373 \\
\hline
\end{tabular}
\end{table*}
\begin{table}[t]
\centering
\caption{Details of our proposed AT-ADD Track 1 dataset.}
\label{tab:t1_train_dev}
\begin{tabular}{lrr}
\hline
\textbf{Model / Condition} & \textbf{Train} & \textbf{Dev} \\
\hline
Real  & 9,999 & 10,000 \\
\hline
ProDiff~\cite{huang2022prodiff} & 1,999 & 1,998 \\
PortaSpeech (normal)~\cite{ren2021portaspeech} & 1,996 & 1,994 \\
DiffSpeech~\cite{liu2022diffsinger} & 1,998 & 1,999 \\
FastSpeech2~\cite{ren2021fastspeech} & 1,998 & 1,998 \\
Kokoro~\cite{nayak2025kokoro} & 2,000 & 1,998 \\
WaveNet~\cite{oord2016wavenet} & 1,999 & 2,000 \\
FastDiff~\cite{huang2022fastdiff} & 1,996 & 1,994 \\
MeloTTS~\cite{zhao2023melotts} & 1,996 & 1,988 \\
CosyVoice~\cite{du2024cosyvoice} & 1,835 & 2,000 \\
Parler-TTS (mini)~\cite{lacombe2024parlertts} & 1,998 & 1,995 \\
GradTTS~\cite{popov2021gradtts} & 1,994 & 1,997 \\
FastPitch~\cite{lancucki2021fastpitch} & 1,998 & 1,997 \\
Tacotron2~\cite{shen2018tacotron2} & 1,994 & 1,998 \\
Glow-TTS~\cite{kim2020glow} & 2,000 & 2,000 \\
WaveGlow~\cite{prenger2019waveglow} & 1,997 & 1,997 \\
MultiBandMelGAN~\cite{kumar2019melgan} & 1,997 & 2,000 \\
Tortoise-TTS~\cite{betker2023scaling} & 1,992 & 1,989 \\
StarGANv2-VC~\cite{li2021starganv2vc} & 1,999 & 1,999 \\
Llasa 1B~\cite{ye2025llasa} & 1,793 & 1,795 \\
Index-TTS~\cite{deng2025indextts} & 1,997 & 1,998 \\
\hline
\textbf{Total} & \textbf{49,575} & \textbf{49,734} \\
\hline
\end{tabular}
\end{table}

\subsection{Track 1: Robust Speech Deepfake Detection}

We construct the AT-ADD Track~1 dataset with predefined training, development, and evaluation splits. The evaluation split is reserved for testing and consists of real speech collected from diverse domains, together with fake speech generated by methods that are unseen in the training and development sets, thereby enabling the evaluation of CM robustness under domain shifts, such as variations in recording devices, acoustic environments, and signal perturbations. Table~\ref{tab:t1_train_dev} summarizes the overall composition of AT-ADD Track~1 dataset. 

\textbf{Real speech in train/dev sets.}
The real speech subset within the training and development sets is comprised of a diverse collection of multilingual utterances. This subset incorporates internal recordings captured across a variety of Recording devices to ensure acoustic diversity, alongside high-quality Chinese speech samples sourced from the AISHELL-3~\cite{shi2020aishell} dataset. To bolster the English portion, samples are integrated from the LibriTTS-R~\cite{koizumi2023libritts} and LJSpeech~\cite{ljspeech} corpora. Furthermore, the dataset’s multilingual breadth is further extended through the inclusion of representative samples from Common Voice~\cite{ardila2020common}, covering a wide array of linguistic contexts.

\textbf{Fake speech in the train/dev sets.}
The fake speech subset covers multiple tasks, including text-to-speech (TTS) and voice conversion (VC). For TTS, the input texts are selected from the real speech data described in the previous subsection and used for synthesis. For VC and one-shot TTS tasks that require reference speaker cloning, we use the same pool of real reference speakers for the training and development sets, while a different pool of real reference speakers is used for the evaluation set to prevent speaker-information leakage. In addition, during cloning, we ensure that the reference speaker and the source speaker are never the same person. 

\textbf{Eval sets.}
The real speech in the evaluation set is collected from both self-recorded data captured with diverse recording devices and additional out-of-domain (OOD) public-source data. The fake speech in the evaluation set is generated by 26 methods that are \emph{unseen} during training. In terms of synthesis paradigms, the fake speech covers several mainstream categories, including vocoder-based, codec-based, and diffusion-based methods. Furthermore, a portion of the fake speech is further replayed to simulate replay attack scenarios, while another portion of both real and fake speech is subjected to signal perturbations to thoroughly evaluate countermeasure robustness. Importantly, neither replay nor signal perturbation changes the original real/fake label of an audio sample; rather, they are regarded as markers for evaluating CM robustness.

\begin{table}[t]
\centering
\caption{AT-ADD Track 2 subsets: \texttt{train}/\texttt{dev} composition across audio types.}
\label{tab:t2_all_subsets}
\begin{tabular}{llrr}
\hline
\textbf{Type} & \textbf{Model / Source} & \textbf{Train} & \textbf{Dev} \\
\hline

\multirow{4}{*}{Sound}
& Real & 9,854 & 4,940 \\
& AudioLDM~\cite{liu2023audioldm} & 9,995 & 4,997 \\
& AudioLDM 2~\cite{liu2024audioldm2} & 10,000 & 5,000 \\
& AudioGen~\cite{kreuk2022audiogen} & 9,991 & 4,992 \\
\cline{2-4}
& \textbf{Total} & \textbf{39,840} & \textbf{19,929} \\

\hline

\multirow{4}{*}{Singing}
& Real & 9,000 & 4,000 \\
& Soft-VITS-SVC~\cite{sovits2023svc} & 9,000 & 4,000 \\
& NeuCoSVC~\cite{sha2024neural} & 9,000 & 4,000 \\
& SeedVC~\cite{liu2024zeroshotvc} & 9,000 & 4,000 \\
\cline{2-4}
& \textbf{Total} & \textbf{36,000} & \textbf{16,000} \\

\hline

\multirow{6}{*}{Music}
& Real & 4,297 & 536 \\
& MusicGen~\cite{copet2024simple} & 4,212 & 1,204 \\
& MusicLDM~\cite{chen2024musicldm} & 4,276 & 1,221 \\
& AudioLDM2~\cite{liu2024audioldm2} & 4,289 & 1,222 \\
& Stable Audio Open~\cite{evans2025stableaudioopen} & 4,292 & 1,223 \\
\cline{2-4}
& \textbf{Total} & \textbf{21,366} & \textbf{5,406} \\

\hline
\end{tabular}
\end{table}

% \begin{table*}[t]
% \centering
% \caption{AT-ADD statistics (number of clips) for Track 1 and Track 2.}
% \label{tab:alldata_protocol}
% \begin{tabular}{c c | ccccc}
% \hline
% \multirow{2}{*}{\textbf{Split}} & 
% \multirow{2}{*}{\textbf{T1}} & 
% \multicolumn{5}{c}{\textbf{T2}} \\
% \cline{3-7}
%  &  & \textbf{Speech} & \textbf{Sound} & \textbf{Singing} & \textbf{Music} & \textbf{Total} \\
% \hline
% Train    & 49,575  & 49,575  & 39,840 & 36,000 & 21,366 & 146,781 \\
% Dev      & 49,734  & 49,734  & 19,929 & 16,000 & 5,406  & 91,069 \\
% Progress & 29,269  & 28,813  & 5,729  & 4,872  & 6,461  & 45,875 \\
% Eval     & 146,346 & 144,078 & 28,593 & 24,332 & 32,370 & 229,373 \\
% \hline
% \end{tabular}
% \end{table*}

\subsection{Track 2: All-Type Audio Deepfake Detection}

In this section, we describe the composition of the proposed AT-ADD Track~2 dataset across four different audio types. Table~\ref{tab:t2_all_subsets} summarizes the overall composition of AT-ADD Track~2 dataset. 

\textbf{Speech.}
We use the same speech training and development sets as in AT-ADD Track~1. However, the evaluation set is simplified by removing the signal perturbation and replay attak introduced in Track~1. As Track~2 targets universal, all-type deepfake detection, this design provides a clean and consistent evaluation protocol that emphasizes cross-type generalization rather than robustness to signal degradations.

\textbf{Sound.}
The sound subset is constructed from the AudioCaps dataset~\cite{kim2019audiocaps}. We first divide the audio samples in AudioCaps, which labeled as real samples, into non-overlapping training, development, and evaluation sets. The synthetic samples in the training and development sets are generated by TTA models conditioned on the corresponding textual descriptions of the real audio. For the evaluation set, the fake samples are generated from the remaining textual descriptions using 4 \emph{unseen} generation methods. In addition, we include OOD real sound samples from other public datasets in the evaluation set to assess the generalization capability of CMs.

\textbf{Singing Voice.}
The singing voice subset is constructed from three source datasets: OpenCpop~\cite{wang2022opencpop}, M4Singer~\cite{zhang2022m4singer}, and KiSing~\cite{shi2024singing}. As in the sound subset, the samples from these three source domains are first divided into non-overlapping training, development, and evaluation sets, which are labeled as the real samples. The fake samples in the training and development sets are generated via singing voice conversion, with strictly non-overlapping source and target singers to avoid identity leakage. The fake samples in the evaluation set are produced by 5 \emph{unseen} deepfake methods, allowing a comprehensive evaluation of cross-model generalization.

\textbf{Music.}
The music subset is derived from the MusicCaps \cite{agostinelli2023musiclm} dataset. We first divide the audio samples in MusicCaps, which labeled as real samples, into non-overlapping training, development, and evaluation sets. The synthetic samples in the training and development sets are generated by TTM models conditioned on the corresponding textual descriptions of the real music. For the evaluation set, the fake samples are generated from the remaining textual descriptions using 4 \emph{unseen} generation methods. In addition, we include OOD real music samples from other public datasets in the evaluation set to assess the generalization capability of CMs.

\section{Baselines}
\label{sec:baseline}

To facilitate fair comparison and lower the entry barrier, we provide a set of official baselines covering conventional CMs, SSL-based CMs, and ALLM-based CMs. These baselines span different modeling paradigms and serve as strong and reproducible starting points for participants.

\begin{table*}[t]
\centering
\caption{Baseline performance (\%) on the AT-ADD benchmarks. \textbf{Best results are highlighted in bold.}}
\label{tab:baseline_results}
\resizebox{\textwidth}{!}{
\begin{tabular}{llccc|ccc|cccc|cccc}
\toprule
\multirow{2}{*}{Type} & \multirow{2}{*}{Model} 
& \multicolumn{3}{c|}{T1} 
& \multicolumn{3}{c|}{T2} 
& \multicolumn{4}{c|}{T2 Progress (by Type)} 
& \multicolumn{4}{c}{T2 Eval (by Type)} \\
\cmidrule(r){3-5} \cmidrule(r){6-8} \cmidrule(r){9-12} \cmidrule(l){13-16}
& 
& Dev & Progress & Eval
& Dev & Progress & Eval
& Speech & Sound & Singing & Music
& Speech & Sound & Singing & Music \\
\midrule

\multirow{2}{*}{Conventional} 
& Spec-ResNet        
& 81.85 & 47.93 & 47.41 
& 57.51 & 53.22 & 53.83 
& 51.08 & 52.92 & 48.41 & 60.48 
& 51.79 & 54.35 & 49.29 & 59.88 \\

& AASIST             
& 94.16 & 60.78 & 60.39 
& 93.63 & 62.38 & 62.21 
& 63.69 & 56.88 & 64.08 & 64.87 
& 63.58 & 56.62 & 63.81 & 64.85 \\

\midrule

\multirow{2}{*}{SSL-based} 
& FT-XLSR-AASIST     
& \textbf{99.70} & \textbf{76.98} & \textbf{76.73} 
& \textbf{98.48} & \textbf{79.25} & \textbf{79.47} 
& \textbf{79.43} & \textbf{66.08} & \textbf{96.33} & \textbf{75.17} 
& \textbf{79.50} & \textbf{66.82} & \textbf{96.30} & \textbf{75.28} \\

& WPT-XLSR-AASIST    
& 96.27 & 73.56 & 73.35 
& 95.00 & 66.59 & 66.68 
& 69.42 & 52.97 & 79.81 & 64.17 
& 69.31 & 53.83 & 79.56 & 64.04 \\

\midrule

\multirow{2}{*}{ALLM-based} 
& Qwen2.5-Omni-3B    
& 93.97 & 68.65 & 68.02 
& 94.44 & 63.42 & 63.23 
& 69.47 & 50.38 & 66.31 & 67.52 
& 68.74 & 50.41 & 65.78 & 67.99 \\

& Qwen2.5-Omni-7B    
& 95.93 & 69.19 & 68.64 
& 94.70 & 61.48 & 61.78 
& 69.89 & 45.28 & 68.04 & 63.77 
& 69.29 & 45.94 & 68.03 & 63.87 \\

\bottomrule
\end{tabular}
}
\end{table*}
\subsection{Baseline Models}
We provide official implementations for all baseline systems used in AT-ADD, including conventional and SSL-based baselines\footnote{\url{https://github.com/xieyuankun/AT-ADD-Baseline}} as well as the ALLM-based baseline\footnote{\url{https://github.com/yangchunmian123/AT-ADD-ALLM-Baseline}}. We next give a brief introduction to these baseline models.

\textbf{Conventional CMs.}
These models follow the traditional pipeline of feature extraction and discriminative classification, representing standard approaches in audio deepfake detection. Although generally weaker than recent SSL-based methods, they serve as important reference systems for evaluating robustness and cross-type generalization.

\begin{itemize}[leftmargin=*]
    \item \textbf{Spec-ResNet}: A spectrogram-based detector with a ResNet backbone~\cite{he2016deep}, representing traditional feature-based CMs. In this baseline, we use only STFT-based spectrograms as input features, without incorporating specialized speech features such as Mel-spectrograms. This design aims to investigate whether simple, generic spectral representations can provide robustness to noise and support cross-type generalization.

    \item \textbf{AASIST~\cite{jung2022aasist}}: A raw waveform-based model employing a sinc convolution front-end~\cite{ravanelli2018speaker} and residual blocks, followed by spectral--temporal attention for classification. This baseline operates directly on raw waveforms without explicit feature engineering, allowing us to study the performance of end-to-end waveform-based detection.
\end{itemize}

\textbf{SSL-based CMs.}
To improve robustness and generalization, we further include models enhanced by SSL representations. These approaches leverage large-scale pre-training and have become the dominant paradigm in modern CM systems, demonstrating strong performance.

\begin{itemize}[leftmargin=*]
    \item \textbf{FT-XLSR-AASIST~\cite{tak2022automatic}}: An enhanced AASIST model using self-supervised representations Wav2Vec2-XLSR\footnote{https://huggingface.co/facebook/wav2vec2-xls-r-300m} as the front-end, providing improved transferability across languages and domains~\cite{phukan2024heterogeneity,pascu24_interspeech}. FT denotes full fine-tuning (FT) of all layers in XLSR and AASIST. To date, it remains a competitive baseline for audio deepfake detection.

    \item \textbf{WPT-XLSR-AASIST~\cite{xie2025detect}}: A strengthened SSL-based baseline incorporating wavelet prompt tuning (WPT) to capture frequency-invariant artifacts. By only optimizing a small number of prompt tokens, this method achieves strong performance with minimal training cost.
\end{itemize}

\textbf{ALLM-based CMs.}
We further provide ALLM-based CMs built on the Qwen audio family.These models take audio (optionally with textual instructions) as input and generate textual outputs, which are adapted into binary real/fake predictions via supervised fine-tuning (SFT). Compared to conventional CMs, ALLM-based approaches leverage large-scale multimodal pre-training and exhibit strong potential for cross-type generalization and unified modeling across heterogeneous audio domains.

\begin{itemize}[leftmargin=*]
    \item \textbf{Qwen2.5-Omni-3B / 7B}\footnote{https://huggingface.co/Qwen/Qwen2.5-Omni-3B}\footnote{https://huggingface.co/Qwen/Qwen2.5-Omni-7B}: Unified multimodal models that support audio inputs and can be adapted for audio deepfake detection via supervised fine-tuning (SFT). Compared to conventional CMs, ALLM-based CMs are capable of producing deterministic predictions and can further provide interpretable reasoning through techniques such as reinforcement learning. Recent studies have demonstrated the superior performance of ALLM-based approaches in the field of ADD~\cite{xie2026interpretable, xue2026unifying, guo2026towards}. Their potential for improving robustness and enabling unified modeling across all audio types makes them a promising direction for further exploration in this challenge.
\end{itemize}

\subsection{Baseline Performance}
Table~\ref{tab:baseline_results} reports the performance of all baselines on both Track~1 and Track~2 under the official evaluation metrics. Notably, for conventional and SSL-based CMs, a unified decision threshold of 0.5 is adopted for real/fake classification.

SSL-based CMs achieve the strongest performance on both tracks. In particular, FT-XLSR-AASIST reaches the best results with $76.73\%$ on Track~1 (Eval) and $79.47\%$ on Track~2 (Eval), consistently outperforming all other baselines. Its performance is also stable across audio types, achieving $79.50\%$ (speech), $66.82\%$ (sound), $96.30\%$ (singing), and $75.28\%$ (music) on the Track~2 evaluation set.

Conventional CMs show significantly lower performance. For example, Spec-ResNet achieves only $47.41\%$ on Track~1 (Eval) and $53.83\%$ on Track~2 (Eval), while AASIST improves to $60.39\%$ and $62.21\%$, respectively, but still remains substantially below SSL-based approaches.

ALLM-based models demonstrate competitive performance despite their general-purpose design. Qwen2.5-Omni-7B achieves $68.64\%$ on Track~1 (Eval) and $61.78\%$ on Track~2 (Eval), while the 3B variant achieves comparable results ($68.02\%$ / $63.23\%$). Notably, ALLM-based models exhibit relatively balanced performance across audio types, suggesting their potential for unified modeling in the all-type ADD setting.

Comparing the two tracks, the overall difficulty of Track~1 and Track~2 is comparable from the baseline results. Track~1 mainly challenges the robustness of models to unseen generators and real-world variations in real audio, while Track~2 highlights performance gaps across audio types (e.g., lower scores on sound and music compared to singing), indicating remaining challenges in achieving fully uniform cross-type generalization.

\section{Rules}

This section outlines the evaluation metrics, and rules for both tracks. Participants are expected to adhere to these rules to ensure a fair and transparent competition.

\subsection{Evaluation Metrics}

The performance of submitted systems is evaluated using the $F1$-score. To ensure fair comparison under class imbalance and across audio types, Macro-$F1$ is adopted with different aggregation strategies for the two tracks.

For a given class $c$, the $F1$-score is defined as the harmonic mean of precision $P_c$ and recall $R_c$:
\begin{equation}
F1_c = \frac{2 \cdot P_c \cdot R_c}{P_c + R_c}
\end{equation}
where
\begin{equation}
P_c = \frac{TP_c}{TP_c + FP_c}, \quad
R_c = \frac{TP_c}{TP_c + FN_c}
\end{equation}
and $TP_c$, $FP_c$, and $FN_c$ denote the numbers of true positives, false positives, and false negatives for class $c$, respectively.

\textbf{Track 1.}
For Track~1 (binary classification: \emph{real} vs.\ \emph{fake}), the official metric is the Macro-$F1$ over the two classes:
\begin{equation}
\text{Macro-}F1_{\text{T1}} = \frac{1}{2}\left(F1_{\text{real}} + F1_{\text{fake}}\right)
\end{equation}
This metric assigns equal importance to both classes, making it robust to class imbalance.

\textbf{Track 2.}
For Track~2, the evaluation accounts for both class balance and audio-type balance. Specifically, for each audio type $t$, we first compute a type-wise Macro-$F1$:
\begin{equation}
\text{Macro-}F1_t = \frac{1}{2}\left(F1_{t,\text{real}} + F1_{t,\text{fake}}\right)
\end{equation}
The final score is then obtained by averaging over all audio types:
\begin{equation}
\text{Macro-}F1_{\text{T2}} = \frac{1}{4} \sum_{t=1}^{4} \text{Macro-}F1_t
\end{equation}
where $t \in \{\text{speech}, \text{sound}, \text{singing}, \text{music}\}$.

Thus, the Track~2 metric enforces a two-level balance: equal weighting across audio types and equal weighting between \emph{real} and \emph{fake} classes within each type.

\subsection{Competition Rules}

The following rules apply to both Track~1 and Track~2 unless otherwise specified.

\paragraph{\textbf{Data Usage.}}
Participants may use only the officially released training and development sets for model training, validation, model selection, and threshold determination. They are free to split the released data for internal training and validation, and may also merge the training and development sets for training. The progress and evaluation sets must not be used in any form for training, fine-tuning, pseudo-labeling, self-training, threshold tuning, or any other kind of model adaptation.

\paragraph{\textbf{External Data.}}
Except for the officially released data, the use of any external labeled or unlabeled audio data is strictly prohibited for training, fine-tuning, distillation, calibration, or pseudo-label construction, including self-generated synthetic data from external generative models or services. Participants must not introduce external datasets related to audio deepfake detection or other closely related authenticity-discrimination tasks, nor may they use models, checkpoints, or feature extractors that have been pre-trained, trained, or fine-tuned on such datasets.

\paragraph{\textbf{Data Augmentation.}}
Data augmentation is allowed only in the form of signal-level perturbation or transformation applied to the officially released data, rather than by introducing external audio data as additional training samples. Allowed augmentation strategies include, but are not limited to, additive noise, reverberation, compression, resampling, and signal-level augmentation methods such as RawBoost. Publicly available augmentation resources, such as MUSAN and RIR libraries, may be used only as augmentation sources and must not be treated as additional supervised training data.

\paragraph{\textbf{Pretrained Models.}}
Publicly available and traceable pretrained models are allowed, including self-supervised learning (SSL) models, audio large language models (ALLMs), multimodal large language models (MLLMs), and other general-purpose pretrained models, provided that their sources can be clearly specified in the final metadata. However, any external models, checkpoints, or feature extractors that have been supervisedly trained or fine-tuned outside AT-ADD for audio deepfake detection or other closely related authenticity classification tasks are strictly prohibited.

\paragraph{\textbf{Fusion and Ensemble.}}
Fusion and ensemble strategies are allowed, including feature-level fusion, score-level fusion, and decision-level fusion. The final submitted system may contain no more than 5 subsystems, and all components must comply with the same data usage rules.

\paragraph{\textbf{Reproducibility.}}
The system corresponding to the final submitted score must be fully automatic and reproducible. The use of opaque closed-source APIs or any other external services that cannot be independently reproduced by the organizers is prohibited. Manual intervention in test set prediction, listening-based correction, or manual annotation is not allowed.

\paragraph{\textbf{Compliance Check.}}
For top-ranked teams, the organizers reserve the right to request a method description, a resource declaration, a model list, and inference code. If any violation of the data usage rules is found, or if a system is determined to be non-reproducible or to involve test-set leakage, the organizers reserve the right to disqualify the submission.

\section{Participation Instructions}

To participate in AT-ADD, teams must first request dataset access by completing the registration form on the Hugging Face dataset page for Track~1\footnote{\url{https://huggingface.co/datasets/xieyuankun/AT-ADD-Track1}} and Track~2\footnote{\url{https://huggingface.co/datasets/xieyuankun/AT-ADD-Track2}}. By participating in the challenge, participants are deemed to have agreed to the corresponding \textit{AT-ADD-Dataset-License}.

After obtaining dataset access, participants are required to register on Codabench for evaluation on Track~1\footnote{\url{https://www.codabench.org/competitions/15477}} and Track~2\footnote{\url{https://www.codabench.org/competitions/15481}}. Each team is allowed to use only \textbf{one} Codabench account, and the email address used for Codabench registration must be consistent with that used during Hugging Face registration.

For submission, participants must upload a \texttt{.zip} file, whose filename can be arbitrary. The compressed file must contain a single file named \texttt{predict.csv}. The required format is:

\begin{verbatim}
name,predict
ATADD_T1_Eval_000001.flac,fake
ATADD_T1_Eval_000002.flac,real
...
\end{verbatim}

Here, \texttt{name} denotes the audio filename, and \texttt{predict} indicates the predicted label, which must be either \texttt{real} or \texttt{fake}.

\section{Challenge Schedule}

The AT-ADD challenge schedule is strictly aligned with the ACM Multimedia 2026 Grand Challenge timeline. The competition consists of a development phase, a final evaluation period, and a technical reporting stage.

\vspace{-1mm}
\begin{table}[htbp]
\centering
\caption{AT-ADD Challenge Important Dates}
\vspace{-3mm}
\label{tab:schedule}
\begin{tabular}{lr}
\toprule
\textbf{Event} & \textbf{Date (2026)} \\ 
\midrule
Release of Data, Baseline Code, and Paper & April 8 \\
Opening of Progress Evaluation Stage & April 8 \\
Opening of Final Evaluation Stage & June 8 \\
Final Leaderboard Freeze & June 15 \\
Metadata and Technical Report Submission & June 17\\
\textbf{Participant Paper Submission Deadline} & \textbf{June 25} \\
Notification of Paper Acceptance & July 16 \\
\textbf{Camera-Ready Paper Submission Deadline} & \textbf{August 6} \\
\bottomrule
\end{tabular}
\end{table}

The challenge schedule is organized into the following key stages:
\begin{itemize}[leftmargin=*]
    \item \textbf{Progress Evaluation Phase (April 8 -- June 11):} Upon the official release of the dataset, baseline code, and challenge paper on April 8, the Progress evaluation stage for all tracks opens on April 8. During this phase, participants can utilize the provided training and development sets to optimize their models and receive real-time feedback via the preliminary leaderboard.
    \item \textbf{Final Evaluation Phase (June 8--June 15):} The final evaluation phase begins on June 8, during which participants in all tracks are required to submit predictions on the full evaluation set. The final leaderboard will be frozen on June 15. The final evaluation dataset will be announced to all registered participants on June 8 via the Codabench mailing system.
    
    \item \textbf{Metadata and Technical Report Submission (June 15--June 17):} After the leaderboard is frozen, each team is required to submit a metadata form describing its submitted system. The detailed format and submission instructions will be sent by email to all teams that have participated in the evaluation phase, and the submission deadline is June 17. In addition to the mandatory metadata, teams may also submit optional technical reports to present their methods in greater detail. These reports will also be displayed on the AT-ADD official website. Although optional, such reports are strongly encouraged, as they help the organizers better understand the proposed methods. Based on the final rankings and submitted materials, three awards will be presented: first place in Track 1, first place in Track 2, and the Best Solution Award. The Best Solution Award will be determined by an expert panel based on algorithmic innovation, practical applicability, interpretability, and reusability.
    \item \textbf{Paper Submission and Review (June 25 -- July 16):} The three award-winning teams will be invited to prepare and submit their papers to the official ACM MM submission system by June 25. These papers will undergo the standard peer-review process of ACM Multimedia, with acceptance notifications scheduled for July 16.
    \item \textbf{Final Camera-Ready Submission (August 6):} Authors of accepted papers are required to submit their final camera-ready versions by August 6 for inclusion in the main conference proceedings.
\end{itemize}

\bibliographystyle{ACM-Reference-Format}
\bibliography{myrefs}

%\appendix
%Lorem ipsum dolor sit amet, consectetur adipiscing elit. Morbi
%malesuada, quam in pulvinar varius, metus nunc fermentum urna, id
%sollicitudin purus odio sit amet enim. Aliquam ullamcorper eu ipsum
%vel mollis. Curabitur quis dictum nisl. Phasellus vel semper risus, et
%lacinia dolor. Integer ultricies commodo sem nec semper.
%Etiam commodo feugiat nisl pulvinar pellentesque. Etiam auctor sodales
%ligula, non varius nibh pulvinar semper. Suspendisse nec lectus non
%ipsum convallis congue hendrerit vitae sapien. Donec at laoreet
%eros. Vivamus non purus placerat, scelerisque diam eu, cursus
%ante. Etiam aliquam tortor auctor efficitur mattis.
%Nam id fermentum dui. Suspendisse sagittis tortor a nulla mollis, in
%pulvinar ex pretium. Sed interdum orci quis metus euismod, et sagittis
%enim maximus. Vestibulum gravida massa ut felis suscipit
%congue. Quisque mattis elit a risus ultrices commodo venenatis eget
%dui. Etiam sagittis eleifend elementum.
%
%Nam interdum magna at lectus dignissim, ac dignissim lorem
%rhoncus. Maecenas eu arcu ac neque placerat aliquam. Nunc pulvinar
%massa et mattis lacinia.

\end{document}